\documentclass[aps,prl,10pt,twocolumn,showpacs,preprintnumbers,amsmath,amssymb,superscriptaddress]{revtex4-1}

\usepackage{listings}

\usepackage[caption=false]{subfig}
\usepackage{lineno}
\usepackage{graphicx}  
\usepackage{dcolumn}   
\usepackage{bm}        
\usepackage{amssymb}   
\usepackage{amsmath}
\usepackage{blkarray, multirow, graphicx, diagbox, color, xcolor, colortbl}
\usepackage{bbm, bbold}
\usepackage{ifthen}
\usepackage[colorlinks, linkcolor = blue, citecolor = blue, filecolor = black, urlcolor = blue]{hyperref}
\usepackage{xkeyval}
\usepackage{moreverb}
\usepackage{rotating}
\usepackage{slashbox}
\usepackage{xspace}
\usepackage{nicefrac}
\usepackage[]{units}
\usepackage[]{natbib}
\usepackage{physics}
\usepackage{braket}
\usepackage{hyphenat}
\usepackage{amsbsy}

\usepackage{ulem}

\begin{document}

\title{Digital Quantum Simulation of Non-Equilibrium Quantum Many-Body Systems}

\author{Benedikt Fauseweh}
\email{fauseweh@lanl.gov}
\affiliation{Theoretical Division, Los Alamos National Laboratory, Los Alamos, New Mexico 87545, USA}

\author{Jian-Xin Zhu}
\email{jxzhu@lanl.gov}
\affiliation{Theoretical Division, Los Alamos National Laboratory, Los Alamos, New Mexico 87545, USA}
\affiliation{Center for Integrated Nanotechnologies, Los Alamos National Laboratory, Los Alamos, New Mexico 87545, USA}

\date{\today}

\begin{abstract}
Digital quantum simulation uses the capabilities of quantum computers to determine the dynamics of quantum systems, which are beyond the computability of modern classical computers. A notoriously challenging task in this field is the description of non-equilibrium dynamics in quantum many-body systems. Here we use the IBM quantum computers to simulate the non-equilibrium dynamics of few spin and fermionic systems. Our results reveal, that with a combination of error mitigation, noise extrapolation and optimized initial state preparation, one can tackle the most important drawbacks of modern quantum devices. The systems we simulate demonstrate the potential for large scale quantum simulations of light-matter interactions in the near future. 
\end{abstract}

\maketitle

Discovering ways to control the dynamical aspects of quantum materials is an important research frontier in modern solid state physics. Advances in non-linear optics and ultrafast spectroscopy allowed to induce non-equilibrium states with new properties on picosecond time scales. In recent years many fascinating phenomena have been observed with this approach, such as light induced superconductivity \cite{Fausti189,Mitrano2016} or ultra fast switching of topological properties \cite{osti_1492730}. These phenomena can lead to desirable electronic and magnetic properties for future devices with potentially groundbreaking applications. 

Theoretical understanding or even prediction of these effects is still scarce, due to computational challenges. The main issue comes from the many interacting particles that participate in such non-equilibrium phenomena, as the classical computational effort to simulate a system of quantum particles scales exponentially in their number $N$.  This is called the curse of dimensionality. 

A possible solution to this problem was proposed by Feynman in 1982 in the form of digital quantum simulation (DQS) \cite{Feynman1982}. It is based on the idea that a quantum system is best simulated by another quantum system. The main ingredient for DQS is a universal quantum computer. A quantum computer uses the effects of quantum mechanics
to execute algorithms, which are substantially faster than their classical counterparts. Quantum complexity theory describes the advantages of these algorithms by providing lower bounds for the time complexity when compared to a classical Turing machine. 
With quantum complexity theory it is possible to prove that the simulation of quantum systems scales only polynomial on a general purpose quantum computer \cite{Lloyd1073}. 

This theoretical superiority has the price of a number of technical challenges \cite{Preskill2018quantumcomputingin}, the biggest one being decoherence of the quantum information, i.e. the decay of the information encoded in a quantum register. 
 In recent years several different approaches were successful in building small scale quantum computers. While first implementations were based on nuclear magnetic resonance in liquids \cite{Cory1634,Gershenfeld350,Vandersypen2001}, recent implementations have been based on superconducting  circuits \cite{Wendin_2017} and trapped ions \cite{doi:10.1063/1.5088164}. 
  However, with the increase of qubits in these quantum devices (up to 72 qubits nowadays), the noise and error rates prohibit high fidelity use of all qubits for fault tolerant quantum computation \cite{PhysRevA.52.R2493,PhysRevLett.77.793,PhysRevA.55.900}.  
The first breakthrough in DQS was based on superconducting circuits \cite{PhysRevX.5.021027,Barends2015,Barends2016}, demonstrating the principle feasibility of the approach.

Recent research in DQS has been focused on evaluating the performance of these Noisy Intermediate-Scale Quantum (NISQ) devices for the purpose of time-dependent quantum simulation \cite{Smith2019, PhysRevLett.121.170501,Zhukov2018,PhysRevA.98.032331,CerveraLierta2018exactisingmodel,Chiesa2019, PhysRevB.101.014411,PhysRevLett.124.230501, PhysRevLett.121.086808,PhysRevA.101.052337}.

{Another major research direction is the simulation of open quantum systems on NISQ devices, to realistically study the interaction between a quantum system and its environment \cite{Wineland2000,Rainer2010,Blatt2011,PhysRevA.83.062317,PhysRevB.95.134501,Wang2018}. This allows to investigate not only systems in the Markovian regime, but also long-lived system-environment correlations. Recently the IBM quantum computers were also used to simulate open quantum systems \cite{Perez2020}. }

 These studies showed that, in its pure form, DQS is still in its infancy on NISQ devices due to the low fidelity and decoherence of the qubits, suggesting that dealing with errors has the highest priority.

{In this paper we tackle the problem of optimized closed quantum system simulation} by incorporating a series of state-of-the-art methods in order to improve the quality of DQS. {The main aim is to mitigate the influence of the noise and environment effects as well as gate imperfections}. The performance of the approach is demonstrated on the IBM quantum computers by simulating non-equilibrium excitations in various {closed} systems. We perform three classes of experiments with increasing complexity to evaluate how different aspects of the simulation, such as error mitigation, zero noise extrapolation and optimal state preparation, can affect the circuit and quality of results on NISQ devices. This combination of error mitigation methods has not been explored before for the purpose of DQS and allows us to estimate when DQS will become a serious competition to classical methods.

\section{Results}

\begin{figure*}
\centering
\includegraphics[width=1.0\textwidth]{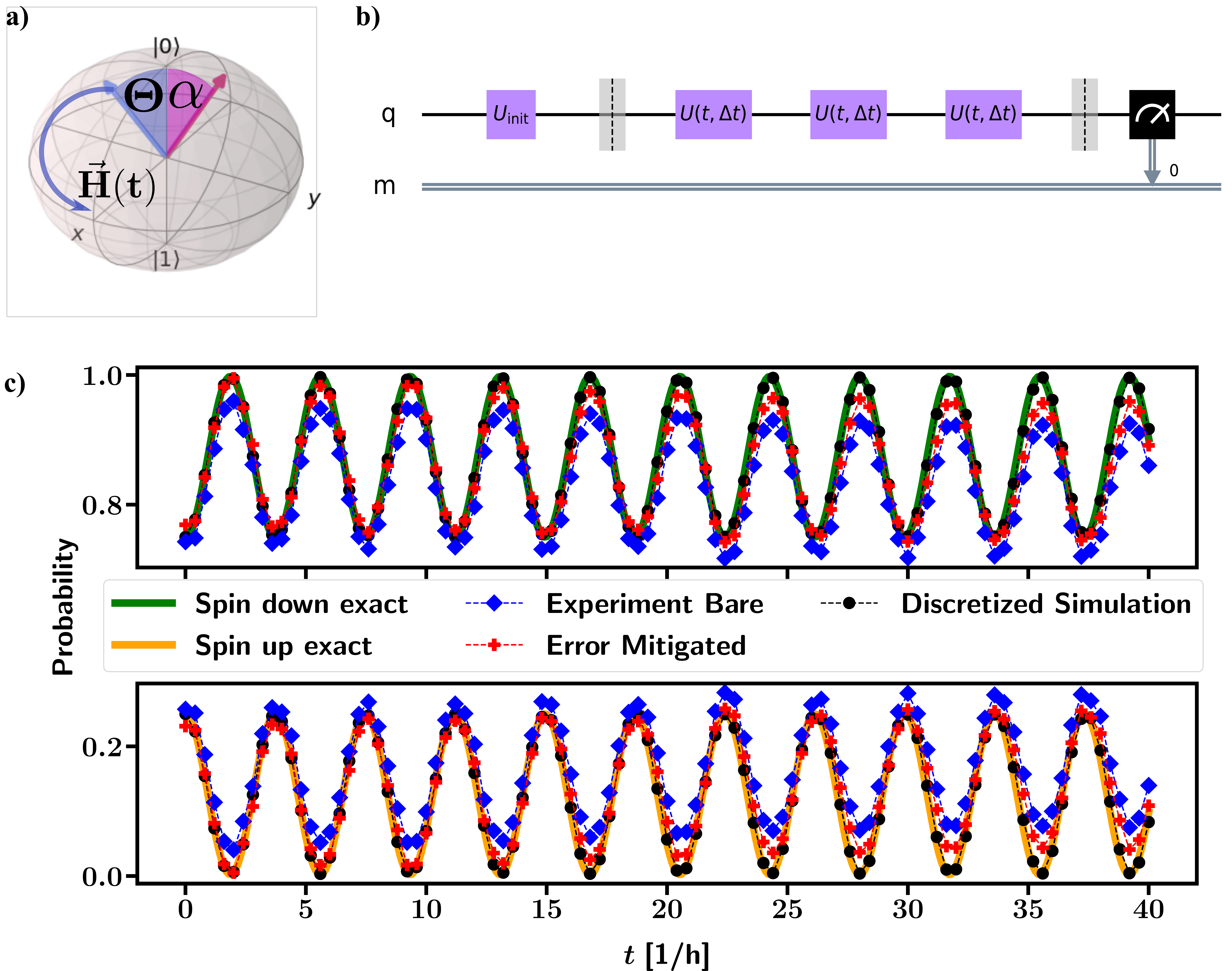}
\caption{{\bf Quantum simulation of the Rabi resonance experiment.} \textbf{a)} Graphical representation of the spin movement on the bloch sphere. The initial state depends on the angle $\alpha$ which rotates the spin between the states $|0\rangle = | \uparrow\rangle$ and $|1\rangle = | \downarrow \rangle$. The magnetic field $\vec{H}(t)$ is constant in magnitude and precesses around the z axis with angle $\Theta$ and angular velocity $\omega$.   \textbf{b)} Quantum circuit to simulate the Rabi experiment on a digitial quantum computer. Initially the state is rotated by an $R_y$ gate by the angle $\alpha$. The unitary time evolution is discretized $t \rightarrow t + \Delta t$ in order to evolve the spin. To reach a time $t=N \Delta t$ we require $N$ single qubit gates $U(t,\Delta t)$. Finally the $z$ component of the spin is measured. \textbf{c)} Results from the quantum simulation for the parameters $H_0 = 1, \Theta = 2, \omega = 1, \alpha = 2 \pi/3$. Comparison between exact results for the probability of measuring either up or down spin, the classically simulated discretized time evolution, the bare experimental results and the error mitigated experimental results. }
\label{fig:f1}
\end{figure*}

\begin{figure*}
\centering
\includegraphics[width=1.0\textwidth]{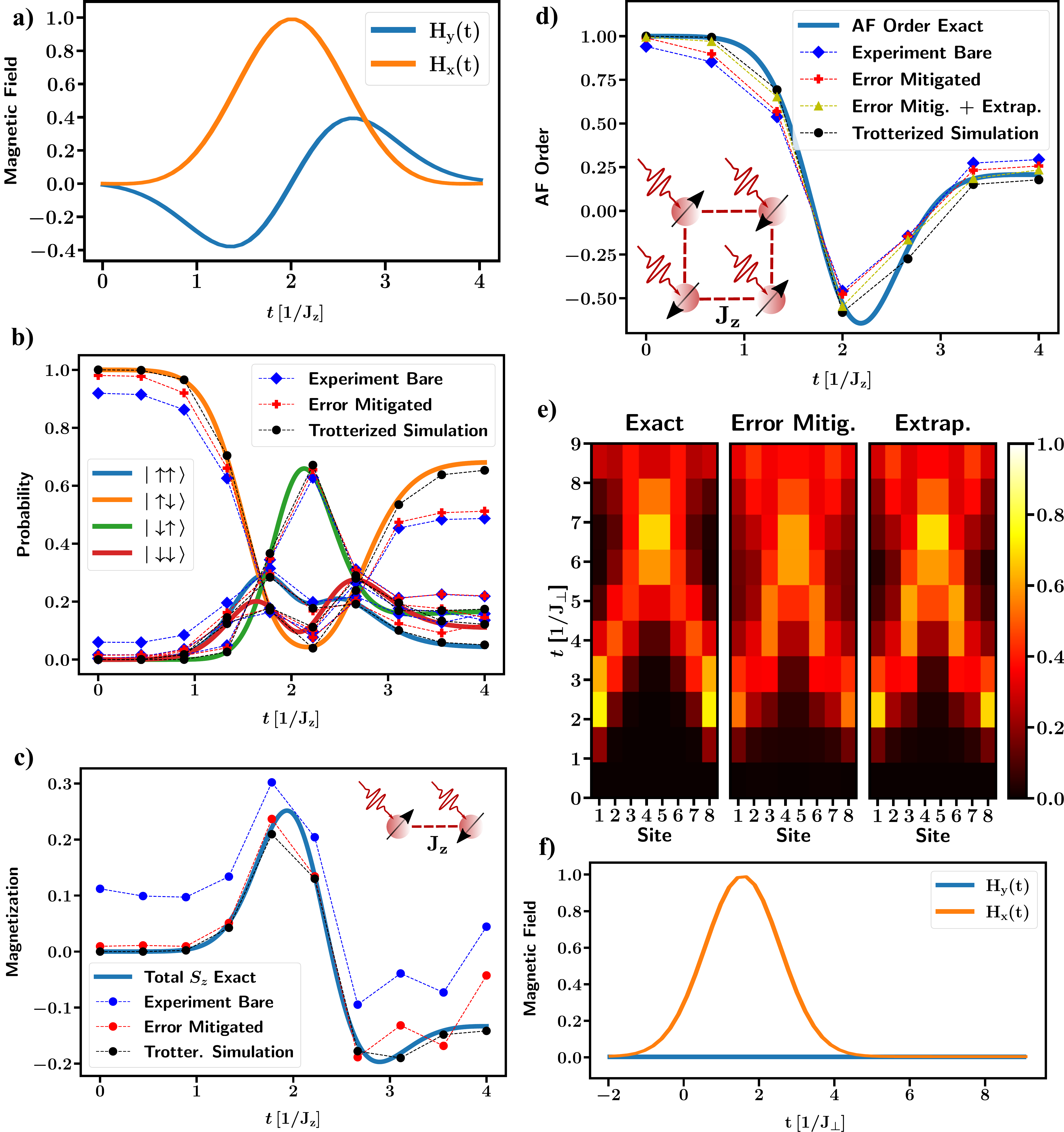}
\caption{{\bf Nonequilibrium quantum simulation of spin models subject to magnetic field pulses.} Only the non-vanishing spin-spin coupling is shown. \textbf{a)} Time dependence of the magnetic field pulse used in \textbf{b)}-\textbf{d)}. The circular polarized pulse is parameterized as $H_x(t) = h_0 \exp(-(t-t_0)/(2 \tau^2)) \cos( \omega (t-t_0))$ and $H_y(t) = h_0 \exp(-(t-t_0)/(2 \tau^2)) \sin( \omega (t-t_0))$ with parameters $h_0 = 2, \omega = 1, \tau=0.7, t_0 = 2$. \textbf{b)} Simulation of a spin dimer in a time-dependent magnetic field. Comparison between the exact time evolution of the states, the bare experimental data, the readout error mitigated data and the classically computed trotterized simulation. \textbf{c)} Time evolution of the total magnetization $S_z(t) = \langle \psi(t) |  S^1_z + S^2_z | \psi(t) \rangle$ derived from \textbf{b)}. \textbf{d)} Simulation of a spin plaquette. Displayed is the staggered magnetization $S_\mathrm{AF} = \sum_{i=1}^4 (-1)^i S^i_z$. Comparison between the bare experimental results, the readout error mitigated results and the combination of readout error mitigation and zero noise extrapolation. \textbf{e)} Time and site resolved magnetization for an edge driven spin chain with $8$ spins. Comparison between the exact results, experimental results after readout error mitigation and the combination of error mitigation and zero noise extrapolation. \textbf{f)} Pulse used for the edge drive in  \textbf{e)}. The pulse is linear polarized and parameterized as $H_x(t) = h_0 \exp(-(t-t_0)/(2 \tau^2)) $ and $H_y(t) =  0$ with parameters $h_0 = \pi/2, \tau=1, t_0 = 1.5$.   }
\label{fig:f2}
\end{figure*}

\begin{figure*}
\centering
\includegraphics[width=1.0\textwidth]{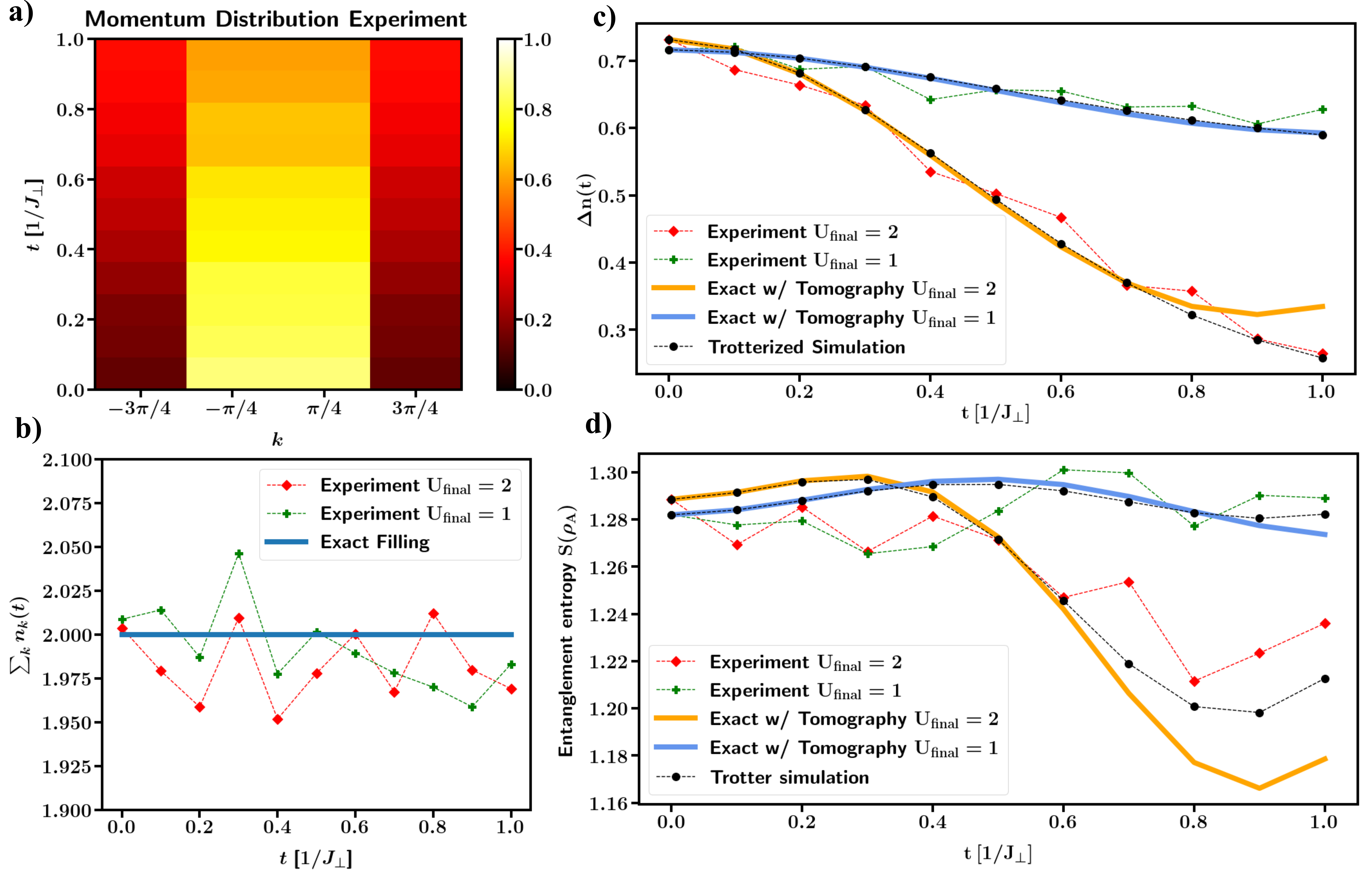}
\caption{{\bf Quantum simulation of an interaction quench in a fermion system.} All experimental data are obtained by combination of optimal state preparation, readout and symmetry error mitigation, zero noise extrapolation and full quantum state tomography after each time step.  \textbf{a)} Time and momentum dependence of the fermionic distribution function on a periodic four site chain after an interaction quench to $U_\mathrm{final} = 2$ at $t=0$. \textbf{b)} Time evolution of the filling factor. Comparison between the exact filling, the weak quench $U_\mathrm{final} = 1$ and the strong quench $U_\mathrm{final} = 2$. \textbf{c)} Time evolution of the jump in the Fermi distribution. Comparison between the experimental results for weak and strong quenches and classical simulations. For the classical simulation the state after initial preparation at $t=0$ was obtained by full state tomography from the quantum computer. It was subsequently evolved by the Schroedinger equation, orange and blue line, and by Trotterization, black dashed line, on a classical computer. \textbf{d)} Entanglement entropy of a bipartition of the system. The classical simulations were obtained with the same method as in \textbf{c)}.  }
\label{fig:f3}
\end{figure*}

Non-equilibrium DQS consists of three fundamental steps. First the initial state is prepared, e.g. an incoming scattering state or the ground state of a Hamiltonian. Second the state is time evolved under the action of time-dependent Hamiltonian. Finally observables are measured after the time evolution. Each of these steps has its own sources of errors, such as imperfect initialization or readout errors. We start by evaluating the single qubit performance of the IBM quantum computers with respect to the simulation of non-equilibrium systems. Specifically we will simulate the spin $S=1/2$  dynamics in a rotating magnetic field $ \vec{H}(t)$,
\begin{align}
\mathcal{H} = \vec{S} \cdot \vec{H}(t).
\end{align} 
This model was first discussed by Rabi in 1937 \cite{PhysRev.51.652} and describes the cyclic behaviour of a two-level quantum system under periodic external perturbation. We identify the computational basis with the spin basis, using the Feynman-Vernon-Hellwarth picture \cite{Feynman1957}. As initial condition we start with a fully polarized spin which is subsequently rotated by an angle $\alpha$ with a $R_y$ gate. The external magnetic field is rotating in the $x$-$y$ plane and has a fixed angle $\Theta$ with respect to the $z$ axis. An overview of the system is shown in Fig. \ref{fig:f1}  a). The quantum circuit for the simulation of the spin dynamics is shown in Fig. \ref{fig:f1}  b). After the initial rotation we discretize the time domain and apply the unitary time evolution $U(t,\Delta t)$ at each time step. Each time evolution can be mapped onto the universal $U3$ gate, up to a global phase. Note that in principle the time evolution for the Rabi model does not require a discretization, as the exact unitary time evolution matrix can be analytically computed. However for simulation of larger systems this approach is not possible, and we therefore want to investigate the performance for the discretized approach already on the level of a single qubit. Finally, after the time evolution, the $z$ component is measured. 
In Fig.  \ref{fig:f1} c) we investigate the results for a representative set of parameters. While the bare experimental results (blue) follow the exact results qualitatively, there is a systematic offset. This offset can be traced back to the readout error of the device. If we assume that this readout error is independent of the gates applied before readout, we can perform a calibration measurement of the single qubit in order to apply readout error mitigation, see Methods section. This approach allows us to significantly reduce the error of our quantum simulation (red). Only in the long time limit we see that decoherence and gate imperfections gradually reduce the accuracy. Note that the error that stems purely from time discretization is negligible (black). 

Next we investigate the non-equilibrium excitation of coupled spin systems. Specifically we investigate pulsed spin dimers and spin plaquettes and an edge driven spin chain with eight spins. The general form of the Hamiltonian we investigate reads
\begin{align}
\mathcal{H} = \sum\limits_{ \langle i,j \rangle} J_\perp \left( S_i^x S_j^x + S_i^y S_j^y \right) + J_z  S_i^z S_j^z + \sum\limits_i \vec{H}_i(t) \vec{S}_i \;,
\end{align}
where $\vec{H}_i(t)$ is a site- and time-dependent magnetic field. We use 2nd order Trotterization to discretize the time evolution of the Hamiltonian, see Methods section. 
The results are shown  in Fig. \ref{fig:f2}. We observe that the reachable time scales before large errors set in have been  reduced almost by an order of magnitude as compared to the single spin case. This effect can be traced back to the two qubit gate errors, specifically the CNOT gate errors, which are typically in the range of $0.5 \%$ up to $3 \%$, depending on the device and the qubits used, see Methods section. Readout error mitigation still significantly reduces the error when compared to the bare experimental results and it has a large effect on global quantities, such as the total magnetization, Fig. \ref{fig:f2} c), or the staggered magnetization, Fig. \ref{fig:f2} d), while the probabilities of the specific states, such as in the spin dimer in  Fig. \ref{fig:f2} b), show larger errors, even after readout error mitigation is made.

To reduce the additional noise coming from the CNOT gates in the DQS circuit we apply zero noise extrapolation \cite{PhysRevLett.119.180509,PhysRevX.7.021050}. The basic idea of this approach is to boost the errors from the two-qubit gates artificially, but in a controlled fashion, in order to map them to the zero noise case. To apply this approach, it is important that the noise itself is time-invariant, i.e. it is invariant under time rescaling. For the superconducting qubits of the IBM quantum devices this was indeed doable~\cite{Kandala2019}.
The extrapolation is implemented together with readout error mitigation for the spin plaquette in Fig.\ \ref{fig:f2} d) and additionally with symmetry error mitigation for a locally driven spin chain in  Fig. \ref{fig:f2} e). Symmetry error mitigation uses the fact, that there is conserved mirror symmetry with respect to the  center of the chain,  even during excitation with an external magnetic field. This mirror symmetry is not necessarily conserved in the quantum circuit, due to gate imperfections. We therefore symmetrize the results from the quantum computer to correct for errors violating this symmetry. 

For the largest system of the $8$-site spin chain, we perform up to $8$ Trotter steps, resulting in a total of $260$ single qubit gates and $126$ entangling gates. Note that for the noise extrapolation, the number of entangling gates is tripled and is therefore $378$ for these measurements.

Individually, these improvements are insufficient to significantly enhance the quality of the experimental results. However, a synergy effect from the combination can lead to quantitative agreement between the experimental results and the exact time evolution. Here the noise extrapolation plays a critical role, as the absolute error for time and site resolved magnetization for the $8$-site spin chain at $t=7/J_\perp$ in Fig. \ref{fig:f2} e) is less than $0.03$ for all sites with extrapolation, while it is up to $0.12$ without extrapolation.

So far we have studied the non-equilibrium dynamics of quantum systems starting from a state which is factorizable. In real materials this situation is typically different. Even at zero temperature quantum fluctuations lead to ground states which are highly entangled. One of the open questions in this context is how such closed many-body system relax after excitation. Here we address this questions for a chain of interacting spinless fermions. 
This model is equivalent to the XXZ chain,
\begin{align}
\label{eq.ll}
\mathcal{H}= \sum\limits_{i \in \mathbb{Z}_4} J_\perp \left(   S_i^x \cdot S_{i+1}^x +  S_i^y \cdot S_{i+1}^y \right)  + U(t) S_i^z \cdot S_{i+1}^z  ,
\end{align}
 by virtue of the Jordan-Wigner transformation \cite{JWTrafo} and it is integrable with the Bethe ansatz \cite{BetheAnsatz}. We use a qubit plaquette to simulate a 
four -site chain with a periodic boundary condition. The system is in its non-interacting ground state for $U(t=0)=0$ and then a sudden quantum quench of the interaction is performed. Such non-adiabatic interaction quenches in Fermi systems are a subject of ongoing research \cite{PhysRevA.80.061602,PhysRevLett.109.126406,PhysRevLett.98.210405,Hamerla_2013,PhysRevLett.115.157201,Essler_2016}.

To initialize the system in its highly entangled ground state we compute the state amplitudes classically and then apply optimized state preparation \cite{2019arXiv190401072I,PhysRevA.93.032318} to reduce the number of noisy CNOT gates. The resulting circuit is given in the Methods section. We apply readout and symmetry error mitigation as well as zero noise extrapolation. After initialization the ground state has a fidelity of $94 \%$ and is pure within statistical error. We then quench the interaction strength in the Hamiltonian and time evolve the state using Trotterization. We compare the effect of two quench strengths  $U_\mathrm{final} = 1$ and  $U_\mathrm{final} = 2$. To compute the time evolution of the momentum distribution we use state tomography after the time evolution and then measure the occupation for each momentum space point. In total our simulation takes on average $68$ single qubit gates, which typically consist of two X90 microwave pulses, and $36$ entangling gates.

The results are given in Fig. \ref{fig:f3}. We observe that although we only have four different momentum space points, the breakdown of the Fermi surface due to interactions is clearly identifiable. Note that the overall occupation seems to be stable with respect to the state preparation and time evolution, as the error stays below $4\%$(Fig. \ref{fig:f3} b)). We trace this observation to the fact that, even in the maximally mixed state, i.e. without any coherence in the system, the total filling is exactly $2$.
In order to compare the results to classical simulations we compute the time evolution using the density matrix obtained from tomography immediately after initial state preparation to assess the errors coming purely from time evolution (blue and orange curves in   Fig. \ref{fig:f3} c) and d)).  The speed of the breakdown and the dependence on the interaction quench is compared in Fig. \ref{fig:f3} c).  There is a good agreement between these classical simulations and the quantum computer experiments. The precision of the simulation allows us to clearly distinguish between the different strengths of the quantum quench.
Note that the maximal time $t_\mathrm{max}= 1 / J_\perp$ that we simulate is sufficient to investigate how the Fermi surface breaks down and how the speed of the breakdown depends on the strength of the interaction. Longer time simulation is possible but shows significant finite size effects, i.e. oscillations in the Fermi jump, which is not representative for the system in the thermodynamic limit.
We also compare the time evolution of the entanglement entropy for a bipartition of the system in Fig. \ref{fig:f3} d).We notice that already at $t=0$ the state is highly entangled due to the fermionic nature of the ground state. The time evolution of the entanglement entropy is only qualitatively captured by the quantum computer, with larger errors when compared to the jump of the Fermi surface. 

\section{Conclusion}
In this paper we have demonstrated the capabilities of modern NISQ devices with respect to simulating non-equilibrium dynamics in quantum system. Using a series of optimizations to improve the accuracy of the simulation circuits, we could significantly outperform DQS in its bare form. While single spin dynamics can be easily captured quantitatively with simple readout error mitigation, multi-spin systems are subjected to much stronger errors, due to the low two-qubit gate fidelities, reducing simulatable time scales when compared to the single spin case.  We therefore applied more refined strategies, such as symmetrization and zero noise extrapolation to simulate pulse driven spin dimers, plaquettes and chains. Finally we focused on highly entangled initial states and combined all previous mitigation techniques with optimal initial state preparation to simulate interaction quench dynamics in small scale Fermi systems. Most importantly our work reaches a precision level that was previously out of reach for similar systems simulated on NISQ devices \cite{Smith2019}.
Our results also go beyond previous simulation of fermionic systems \cite{Barends2015}, as our initial state is already a highly entangled Fermi sea, being more realistic for the description of real materials.

Our work highlights the importance of sophisticated methods in order to simulate non-equilibrium dynamics on modern quantum processors.  So far only small time scales and system sizes can be reached, due to noise and gate imperfections. Recent improvements of the IBM quantum devices \cite{jurcevic2020demonstration}, such as shorter two-qubit gates and dynamic decoupling sequences, can be combined with our approach as well. With further technological improvements we can reach system sizes which are not tractable by classical computers, making DQS a possible candidate for the demonstration of quantum supremacy. NISQ devices available today have already reached the number of required qubits ($ \sim 100$) to outperform classical computers, therefore high priority should be on increasing the fidelity of the devices. Based on our results we estimate, that with another order of magnitude improvement to the fidelity of the two-qubit gates and an increased connectivity of the qubit topology, DQS on real quantum devices will outperform classical algorithms. This is especially interesting for two-dimensional systems, where tensor-network approaches hit a barrier, due to the area law of entanglement entropy.

Once these problems have been solved, many opportunities of applications will emerge for DQS, such as real-time dynamics in strongly correlated systems, quantum chemistry, nuclear physics, quantum chromo dynamics and relativistic quantum field theory. Advances in these fields can have a wide influence on the development of future applications, such as drug discovery, chemical reaction modeling and quantum materials design.

\section{Methods}

\begin{figure*}[h]
\centering
\includegraphics[width=1.0\textwidth]{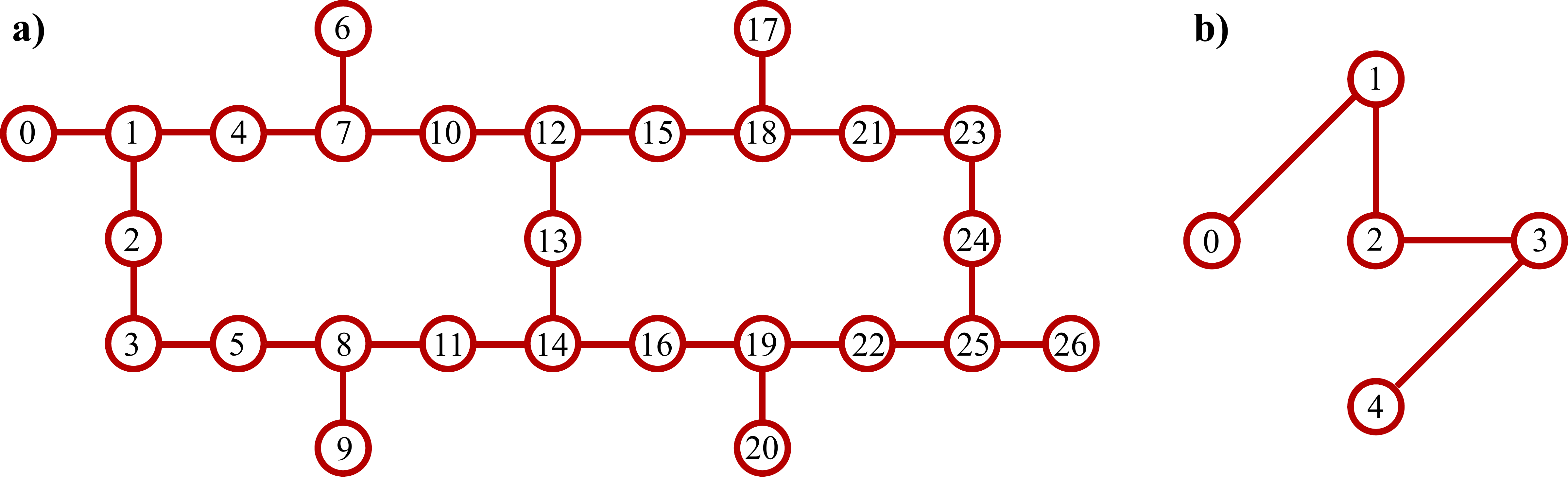}
\caption{{\bf Architecture of the IBM quantum devices used in the paper.} a) Architecture of the ibmq\_montreal and ibmq\_toronto devices. b) Architecture of the ibmq\_bogota device. Lines show the possible CNOT gates on these devices. }
\label{fig:S1}
\end{figure*}

\begin{table*}[h]
\centering
\begin{tabular}{|c|c|c|c|c|} 
\hline
Qubit & $t_1$ [$\mu$s] & $t_2$ [$\mu$s] & U$3$ error rate & readout error rate \\
\hline
14 & $135$ & $221$ & $3.8 \cdot 10^{-4}$ & $1.4 \cdot 10^{-2}$ \\
\hline
\end{tabular}
\caption{Device calibration for ibmq\_toronto device used in Fig. \ref{fig:f1} a). Experiment executed on July 15th 2020.}
\label{tab_rabi}
\end{table*}

\begin{table*}[h]
\centering
\begin{tabular}{|c|c|c|c|c|c|} 
\hline
Qubit & $t_1$ [$\mu s$] & $t_2$ [$\mu s$] & U$3$ error rate & CNOT error rate & readout error rate \\
\hline
19 & $93$ & $75$ & $5.9 \cdot 10^{-4}$ & $19 \rightarrow 20$: $6.7 \cdot 10^{-3}$ & $1.6 \cdot 10^{-2}$ \\
20 & $88$ & $130$ & $4.3 \cdot 10^{-4}$ & $20 \rightarrow 19$: $6.7 \cdot 10^{-3}$ & $3.7 \cdot 10^{-2}$ \\
\hline
\end{tabular}
\caption{Device calibration for ibmq\_montreal device used in Fig. \ref{fig:f2} b) and c). Experiment executed on August 1st 2020.}
\label{tab_dimer}
\end{table*}

\begin{table*}[h]
\centering
\begin{tabular}{|c|c|c|c|c|c|} 
\hline
Qubit & $t_1$ [$\mu s$] & $t_2$ [$\mu s$] & U$3$ error rate & CNOT error rate & readout error rate \\
\hline
8 & $72$ & $135$ & $ 1.0	  \cdot 10^{-3}$ & $ 8 \rightarrow 11$: $ 6.8 \cdot 10^{-3}$ & $ 2.6 \cdot 10^{-2}$ \\
11 & $94$ & $142$ & $ 4.1 \cdot 10^{-4}$ & $11 \rightarrow  8$: $ 6.8 \cdot 10^{-3}$ ,  $11 \rightarrow  14$: $ 1.8  \cdot 10^{-2}$ & $ 1.7 \cdot 10^{-2}$ \\
14 & $89$ & $134$ & $ 4.8 \cdot 10^{-4}$ & $14  \rightarrow 11  $: $ 1.8 \cdot 10^{-2}$  ,  $14 \rightarrow  16$: $ 6.4  \cdot 10^{-3}$ & $ 1.4 \cdot 10^{-2}$ \\
16 & $85$ & $124$ & $ 4.6  \cdot 10^{-4}$ & $ 16 \rightarrow 14  $: $ 6.4 \cdot 10^{-3}$ & $ 1.9 \cdot 10^{-2}$ \\
\hline
\end{tabular}
\caption{Device calibration for ibmq\_toronto device used in Fig. \ref{fig:f2} d). Experiment executed on September 4th 2020.}
\label{tab_plaquette}
\end{table*}

\begin{table*}[h]
\centering
\begin{tabular}{|c|c|c|c|c|c|} 
\hline
Qubit & $t_1$ [$\mu s$] & $t_2$ [$\mu s$] & U$3$ error rate & CNOT error rate & readout error rate \\
\hline
4 & $126$ & $141$ & $  4.2  \cdot 10^{-4}$ & $  4 \rightarrow 1$: $ 1.4  \cdot 10^{-2}$ & $  1.3 \cdot 10^{-2}$ \\
1 & $159$ & $13$ & $  4.7 \cdot 10^{-4}$ & $  1 \rightarrow  4  $: $ 1.4  \cdot 10^{-2}$ ,  $  1 \rightarrow  2 $: $  9.4  \cdot 10^{-3}$ & $ 3.6  \cdot 10^{-2}$ \\
2 & $77$ & $135$ & $ 4.1  \cdot 10^{-4}$ & $  2 \rightarrow  1 $: $  9.4 \cdot 10^{-3}$ ,  $ 2 \rightarrow  3 $: $  1.3  \cdot 10^{-2}$ & $ 1.5  \cdot 10^{-2}$ \\
3 & $110$ & $92$ & $ 4.5  \cdot 10^{-4}$ & $ 3 \rightarrow  2 $: $  1.3 \cdot 10^{-2}$ ,  $ 3 \rightarrow   5$: $  6.5  \cdot 10^{-3}$ & $  1.2 \cdot 10^{-2}$ \\
5 & $142$ & $153$ & $  4.3 \cdot 10^{-4}$ & $ 5 \rightarrow 3  $: $  6.5  \cdot 10^{-3}$ ,  $  5 \rightarrow  8 $: $  6.4  \cdot 10^{-3}$ & $  9.4 \cdot 10^{-3}$ \\
8 & $94$ & $105$ & $ 5.0  \cdot 10^{-4}$ & $ 8 \rightarrow 5  $: $  6.4 \cdot 10^{-3}$ ,  $ 8  \rightarrow  11 $: $ 7.5  \cdot 10^{-3}$ & $  1.0 \cdot 10^{-2}$ \\
11 & $116$ & $47$ & $ 7.7  \cdot 10^{-4}$ & $ 11  \rightarrow 8  $: $ 7.5  \cdot 10^{-3}$ ,  $ 11 \rightarrow   14 $: $  7.2   \cdot 10^{-3}$ & $  2.2 \cdot 10^{-2}$ \\
14 & $138$ & $ 134$ & $ 9.1   \cdot 10^{-4}$ & $ 14  \rightarrow 11 $: $ 7.2  \cdot 10^{-3}$ & $  2.0 \cdot 10^{-2}$ \\
\hline
\end{tabular}
\caption{Device calibration for ibmq\_montreal device used in Fig. \ref{fig:f2} e). Experiment executed on September 4th 2020.}
\label{tab_chain}
\end{table*}

\begin{table*}[h]
\centering
\begin{tabular}{|c|c|c|c|c|c|} 
\hline
Qubit & $t_1$ [$\mu s$] & $t_2$ [$\mu s$] & U$3$ error rate & CNOT error rate & readout error rate \\
\hline
1 & $163$ & $146$ & $ 3.8  \cdot 10^{-4}$ & $ 1 \rightarrow 2$: $  6.6 \cdot 10^{-3}$ & $5.8   \cdot 10^{-2}$ \\
2 & $153$ & $235$ & $ 3.3  \cdot 10^{-4}$ & $2 \rightarrow  1$: $ 6.6  \cdot 10^{-3}$ ,  $2 \rightarrow  3$: $  9.5   \cdot 10^{-3}$ & $ 2.4 \cdot 10^{-2}$ \\
3 & $145$ & $225$ & $  4.5 \cdot 10^{-4}$ & $3  \rightarrow 2  $: $ 9.5  \cdot 10^{-3}$  ,  $3 \rightarrow  4$: $  7.0  \cdot 10^{-3}$ & $ 2.8 \cdot 10^{-2}$ \\
4 & $108$ & $172$ & $  3.6 \cdot 10^{-4}$ & $ 4 \rightarrow 3  $: $ 7.0  \cdot 10^{-3}$ & $  2.4 \cdot 10^{-2}$ \\
\hline
\end{tabular}
\caption{Device calibration for ibmq\_bogota device used in Fig. \ref{fig:f3}. Experiment executed on September 4th 2020.}
\label{tab_fermi}
\end{table*}

\textit{IBM Quantum Experience devices and qubits.} To compute the time evolution we used three different devices from the IBM Quantum Experience. For the experiments shown in Figs.\ \ref{fig:f1} and \ref{fig:f2} we used the 27 qubit devices ibmq\_toronto and ibmq\_montreal. They have an identical architecture, which is shown in Fig.\ \ref{fig:S1} a). For the calculations on the fermionic system we used the 5 qubit device ibmq\_bogota, shown in Fig.\ \ref{fig:S1} b). Note that these devices are recalibrated on a daily basis, which may change the characteristics of the gate error rates as well as the decoherence times. We have summarized the calibration of the qubits used in the experiments in the Tables \ref{tab_rabi}, \ref{tab_dimer}, \ref{tab_plaquette}, \ref{tab_chain} and \ref{tab_fermi}. All results from quantum devices were averaged over 8192 samples, such that the statistical error is negligible.

\textit{Readout error mitigation.} To reduce the error coming from readout, we applied a readout error mitigation technique \cite{Qiskit-Textbook, PhysRevLett.108.057002}. Two primary assumptions are necessary for this approach: 1) The error coming from readout is due to classical noise and 2) that noise is independent of the quantum gates applied to the system beforehand. In recent study it was shown that classical noise is indeed the dominant noise on the IBM quantum devices \cite{Maciejewski2020mitigationofreadout}. We thus
 executed a calibration experiment before each time evolution experiment in order to characterize the device for each of the $2^N$ basis states. Although this approach scales exponentially in the system size, this can be overcome by tensored error mitigation, assuming further that the error by noise is local and correlates only a subset of qubits. After the calibration experiment, we arranged the results of each calibration experiment in a $2^N \times 2^N$ matrix $\Lambda$,
\begin{align}
\Lambda_{ij} = p(i, j)
\end{align}
where $p(i,j)$ is the probability of preparing state $i$ and measuring state $j$.
If the single gate errors are small compared to the readout noise this matrix perfectly maps the ideal results to the experimental results by means of a classical post processing of the statistics
\begin{align}
p_\mathrm{exp} =  \Lambda p_\mathrm{ideal} .
\end{align}
Thus to obtain the ideal results we applied the inverse of $\Lambda$ onto the experimental results of our time evolution. Note that the inverse might not be well defined, due to strong noise. In this case the Moore–Penrose pseudoinverse was applied to obtain a least-square solution. 

\textit{Time evolution.} To time evolve the quantum states after initialization, we used symmetric Trotterization to decompose the time evolution operator.
\begin{widetext}
\begin{equation*}
\begin{aligned}
U(t, \Delta t) &= T \exp \left[  -i \int\limits_{t}^{t+\Delta t} H(\tau) \mathrm{d} \tau  \right] \\
&=   \exp \left[ -i \frac{\Delta t}{2}  H_A \left(t+\frac{\Delta t}{2} \right) \right] \exp \left[ -i \Delta t H_B \left( t+\frac{\Delta t}{2} \right) \right]  \exp \left[ - i \frac{\Delta t}{2} H_A \left( t+\frac{\Delta t}{2} \right) \right] + \mathcal{O}(\Delta t^3) \;,
\end{aligned}
\end{equation*}
\end{widetext}
where $H_A$ and $H_B$ are two in general non-commuting parts of the Hamiltonian. Note that this approach can be generalized to more non-comuting parts as well \cite{Hatano2005}.
In some cases we also used a lower order formula if the main source of error comes from gate imperfections and not from Trotterization. 
In the Hamiltonians we are simulating the interactions are strongly localized. Therefore, we need only one and two-qubit gates to fully implement the necessary time evolution gates. For details on the implementation of these gates we refer to the literature \cite{Smith2019}.

\textit{Zero noise extrapolation.} In order to perform zero noise extrapolation we concentrate on the two qubit gates and assume that the single qubit gates have negligible error rates in comparison. This is supported by the observation that the two qubit error rates are typically one order of magnitude larger than the single qubit gates on the IBM quantum devices, see above. 
Zero noise extrapolation is based on the idea of Richardson's deferred approach to the limit. Suppose we want to measure an expectation value of an observable $E$. Then this expectation value can be expanded as a power series in a noise parameter $\lambda$,
\begin{equation}
E(\lambda) = E_0 + \sum\limits_{k=1}^{\infty} a_k \lambda^k  \;,
\end{equation}
where $E_0$ is the zero noise limit. If we can obtain several expectation values at different noise values $E(\lambda_1), E(\lambda_2), \dots$ it is possible to cancel out the leading contribution in the power series to obtain a better estimate of the zero noise term. To obtain these different noise values it is possible to scale the whole quantum circuit by a factor $c$. In Ref.\ \cite{Kandala2019} this approach was applied to the single as well as to the two qubit gates, by rescaling the duration of the microwave pulses. This rescaling requires a precise recalibration of the two qubit gates due to non-linearities in the amplitude dependence. 
Here we use a simpler scheme by concentrating only on the CNOT gates and fix the stretching factor to $c=3$. Specifically we set up the quantum circuits for $c=1$ and then replace every CNOT gate but three times the same CNOT gate. This approach allows us to boost the noise from the CNOT gate and at the same time perform the same computation, as in theory three repeated CNOTs are equivalent to a single CNOT. We then use the readout of the extended circuit to perform a linear extrapolation of the expectation value, assuming that the single qubit gates have no error and do not contribute to the noise. 

\textit{Fermionic model.} In order to simulate the breakdown of the Fermi surface in Fig.\ \ref{fig:f3} we are simulating the XXZ model on a plaquette as given in Eq. \eqref{eq.ll}. Using the Jordan Wigner transformation \cite{JWTrafo} we obtain an equivalent fermionic model,
\begin{widetext}
\begin{equation*}
H_\text{\tiny LL} =  \frac{J_\perp}{2} \sum\limits_{i = 1}^3 \left( c^\dagger_i c^{\phantom \dagger}_{i+1} + \text{h.c.} \right) +  \frac{J_\perp}{2} (-1)^{N-1)}  \left( c^\dagger_4 c^{\phantom \dagger}_{1} + \text{h.c.} \right) + U \sum\limits_{i \in \mathbb{Z}_4} \left( c^\dagger_i c^{\phantom \dagger}_{i} - \frac{1}{2} \right) \left(  c^\dagger_{i+1} c^{\phantom \dagger}_{i+1} - \frac{1}{2} \right)   \;,
\end{equation*}
\end{widetext}
with $U= J_\perp \Delta$ and $N$ the total number of fermions. Note that we use periodic boundary conditions for the spin model. In this case the ground state for $U=0$ is in the sector with $N = 2$ fermions. Therefore the fermionic model has anti-periodic boundary conditions, which results in the $k$ values $-3\pi/2, -\pi/2, \pi/2, 3\pi/2$. Even after the quantum quench the fermionic Hamiltonian commutes with the total particle number operator and therefore we will stay in the $N = 2$ sector during the time evolution, also seen in Fig.\ \ref{fig:f3} b). 

\begin{figure*}
\centering
\includegraphics[width=1.0\textwidth]{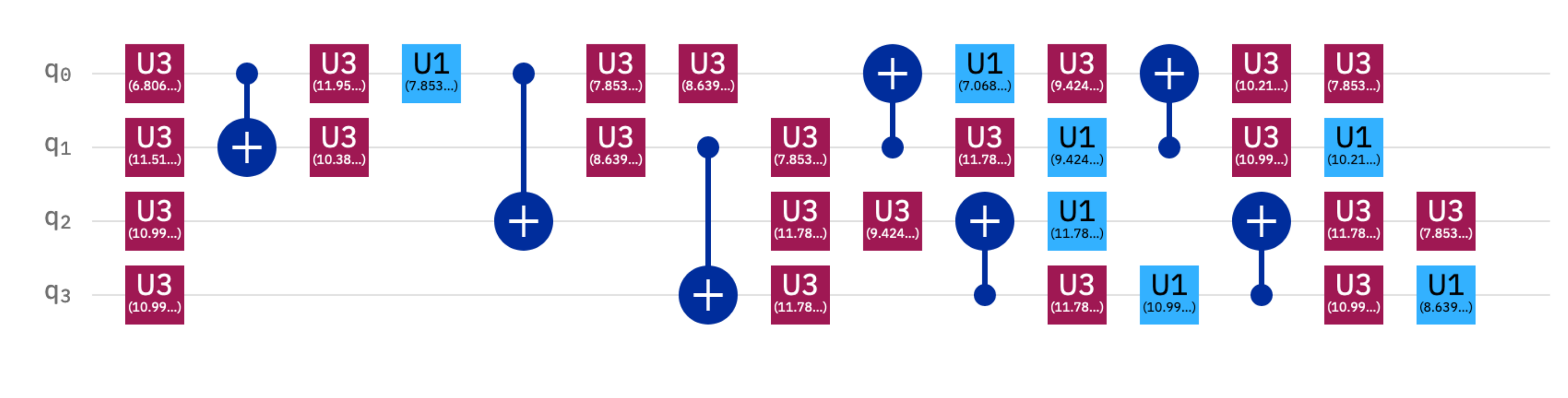}
\caption{{\bf Circuit for optimized state preparation.} Ground state preparation used in Fig.\ \ref{fig:f3}. }
\label{fig:S2}
\end{figure*}

\textit{Optimal state preparation.} For the simulation of the interaction quench in Fig. \ref{fig:f3} we require a highly entangled initial state. Although in theory every state can be constructed on a universal quantum computer \cite{PhysRevA.52.3457}, recent progress was on minimizing the number of required CNOT gates in order to increase the fidelity of the state preparation \cite{PhysRevA.83.032302,PhysRevA.93.032318}. Here we use the universal Q compiler \cite{2019arXiv190401072I} to compute an optimized circuit for the preparation of the fermionic ground state. The result in QASM code is given in Listing 1 and shown in Fig.\ \ref{fig:S2}.

\textit{State tomography and entanglement entropy.} In order to compute the momentum distribution and the entanglement entropy in Fig.\ \ref{fig:f3}, we used full state tomography to deduce the state after each time step. This requires $3^N$ circuits to measure the state in the $X$, $Y$ and $Z$ basis.  Due to noise during the quantum computation, the state is not necessarily a pure state. We used a fitter implemented in the QISKIT package \cite{Qiskit-Textbook} to compute the density matrix $\rho(t)$ based on a method introduced in Ref.~ \cite{PhysRevLett.108.070502}. 
We then used this density matrix to deduce the momentum distribution
\begin{align}
n_k(t) = \frac{1}{N_s} \sum\limits_{i,j} \langle  c^\dagger_i c^{\phantom \dagger}_{j} \rangle (t) e^{ i k (r_i - r_j) } \;,
\end{align}
where $N_s = 4$ is the number of sites and $r_i$ is the position of site $i$. The expectation value $\langle \dots \rangle$ is with respect to the density matrix $\rho(t)$.
The Von Neumann entanglement entropy is computed as
\begin{align}
S(\rho_\mathrm{A})(t) = - \mathrm{Tr} \left[ \rho_\mathrm{A}(t) \log \rho_\mathrm{A}(t) \right] \;,
\end{align}
where $ \rho_\mathrm{A}(t) = \mathrm{Tr}_\mathrm{B} (\rho(t))$ is the reduced density matrix of subsystem A. Here we used a bipartition of the system to define the subsystems A and B. Although this way of computing the entanglement entropy does not scale to large system sizes, there are demonstrations on how to measure fermionic entanglement on quantum computers using an alternate approach \cite{PhysRevA.99.062309}.

\begin{figure*}
\centering
\begin{minipage}{\linewidth}
\begin{lstlisting}[label = listing1, caption = {QASM code for optimal state preparation of the fermionic ground state}]
include "qelib1.inc";
qreg q[4];
u3(6.806784082778, 0, 0) q[0];
u3(11.519173063162, -pi/2, pi/2) q[1];
cx q[0], q[1];
u3(11.950890905689, 0, 0) q[0];
u3(10.380094578894, 0, 0) q[1];
u1(7.853981633974) q[0];
u3(10.995574287564, -pi/2, pi/2) q[2];
cx q[0], q[2];
u3(8.639379797372, -pi/2, pi/2) q[1];
u3(10.995574287564, -pi/2, pi/2) q[3];
cx q[1], q[3];
u3(11.780972450962, 0, 0) q[3];
u3(11.780972450962, 0, 0) q[2];
u3(9.424777960769, -pi/2, pi/2) q[2];
cx q[3], q[2];
u3(11.780972450962, -pi/2, pi/2) q[3];
u1(10.995574287564) q[3];
u1(11.780972450962) q[2];
cx q[3], q[2];
u3(10.995574287564, 0, 0) q[3];
u1(8.639379797372) q[3];
u3(11.780972450962, 0, 0) q[2];
u3(7.853981633974, -pi/2, pi/2) q[2];
u3(7.853981633974, -pi/2, pi/2) q[0];
u3(7.853981633974, 0, 0) q[1];
u3(8.639379797372, 0, 0) q[0];
cx q[1], q[0];
u3(11.780972450962, 0, 0) q[1];
u1(9.424777960769) q[1];
u1(7.068583470577) q[0];
u3(9.424777960769, -pi/2, pi/2) q[0];
cx q[1], q[0];
u3(10.995574287564, 0, 0) q[1];
u1(10.210176124167) q[1];
u3(10.210176124167, 0, 0) q[0];
u3(7.853981633974, -pi/2, pi/2) q[0];
\end{lstlisting}
\end{minipage}
\end{figure*}

\section*{Acknowledgments}

We thank Christopher Lane, Zhao Huang, and Andrew Sornborger for useful discussions.
This work was carried out under the auspices of  the  U.S.  Department  of  Energy  (DOE)  National Nuclear  Security  Administration  under  Contract  No. 89233218CNA000001. It  was supported by the LANL LDRD Program (B.F.) and NNSA ASC Program (J.-X. Z.). This research used resources provided by the LANL Institutional Computing Program.

\clearpage

\normalem

\end{document}